\documentclass[preprint,12pt]{elsarticle}

\usepackage{amssymb}
\usepackage{amsmath}
\usepackage{graphicx}
\journal{the arXiv.org}

\begin{document}

\begin{frontmatter}

\title{Application of Boltzmann equation to investigate kinetic processes of meson and baryon production in the quark-gluon plasma}
\author{Yuriy Ostapov}
\address {Institute of Mathematics, Kiev 01024, Ukraine.
E-mail: yugo.ost@gmail.com }

\begin{abstract}
We consider the meson and baryon production in the quark-gluon plasma as applied to the early Universe and  ultrarelativistic heavy ion  collisions  in collider.
The essential feature of our investigation is to take into account a potential barrier for quarks under  meson and baryon production.
We have used the Boltzmann equation to describe kinetic  processes of quark fusion.
We have derived analytic expressions for solutions of kinetic equations.
This permits to find the dependence of reaction rate  on  a time, temperature, mass and charge of quarks as well as on a string tension connected with the confining potential. 
Next we have  considered in detail the meson and baryon production for some concrete kinetic processes in collisions of ultrarelativistic heavy ions.
Exactly the same results  will be as well true for processes in  expanding Universe.

\end{abstract}

\end{frontmatter}

\section{Introduction}

Investigating the quark-gluon plasma (QGP) is an actual problem of modern physics since one is connected with processes 
in the early Universe and with collisions of ultrarelativistic heavy ions in colliders.
There are three basic directions to study QGP:

\begin{itemize}
\item calculating thermodynamic quantities;
\item phase transitions;
\item kinetic processes.
\end{itemize}

There are monographs devoted to thermodynamics and phase transitions for QGP  \cite {vogt, kohs}.
Quantum chromodynamics for finite temperature permits to study  some thermodynamic problems  of QGP \cite { kap, kohs}.
Many tasks connected with QGP are solved by the  lattice gauge theory \cite {kap, vogt, kohs}.
General questions of QGP kinetics are considered in some books \cite { let, kohs}.
The works  \cite { kuz, let}  are devoted to the Boltzmann equation based on a momentum distribution.
  
As applied to collisions of ultrarelativistic heavy ions the production of mesons and baryons  is studied in \cite {bla, chaud, emel, grish,  vogt, wong, kohs}.
Experimental data  and theoretical investigations have been summarized in   articles \cite { shan, xin, hai, pet, tiw}, et al.
Some works are devoted to the quarkonium as the bound state of quark-antiquark pair\cite { vogt, kohs}. 
As a rule, the production of hadrons is considered from the viewpoint of the parton model \cite {clos}, Ch.18.

Results of researches  concerning  collisions of ultrarelativistic heavy ions have been transferred on analogous processes in the early Universe  \cite {kap, kohs}.

Our investigation permits to find explicit expressions for the rate of  kinetic processes based on the Boltzmann equation for QGP.
To determine the cross-section for quark fusion, we represent hadrons as potential wells.
To describe processes of tunnelling for such wells,  we use the quasi-classical approximation  in quantum mechanics.
We have derived explicit mathematical expressions for solutions of kinetic equations.
Next we have considered  concrete processes of meson and baryon production.
In particular, we have found numerical kinetic  relations for mesons $\pi, J/\psi$ and baryons $p, {\Sigma}^+$.
We have shown as well that these results will be true  for expanding Universe.

\section{Quark-gluon plasma}

 According to quantum statistics the Fermi-Dirac distribution for free quarks (i.e.,  before fusion) is given by the expression:

\begin{equation}
   f(p) = \frac {g}{(2 \pi)^3} \frac {1}{ exp( \frac {E(p)-\mu) } {T }) +1} ,                                                          
\end{equation}
where $g=2$, $p$ and $m$ are a momentum and mass of quark,  $E(p)$ is a quark energy ($E(p) =\sqrt {p^2 +m^2}$), $T$ is a temperature, $\mu$ is a chemical potential of quark.

Using the distribution (1), we shall derive the equilibrium density of light quarks under the condition  $p>>m$ and $\mu=0$:

 \begin{eqnarray}
n_{eq}=  4\pi g \int    d p\, p^2   f(p) \approx 4 \pi g \int_{0}^{\infty}   d E \, E^2   f(E) \nonumber \\
=g \int_{0}^{\infty}\frac {E^2 dE} {2 {\pi}^2 (exp (E/T) +1)} = \frac {3 g \zeta (3) T^3}{4 {\pi}^2 }.
\end{eqnarray}
In the  expression (2) we use the  value of integral from \cite {lif} \S 58 ($\zeta (3) \approx 1.2$).

Consider the  phase transition for QGP \cite { vogt, wong, kohs}. Using the QCD partition function, one can find that the temperature of such transition $T_{c}$
is    180-200  MeV. To get this result, it is necessary to  equate   a pressure and  temperature  for QGP and a hadron gas. 
For example, for mesons

\begin{equation}
   P_{c} =   P_{meson} = P_{QGP},                                                         
\end{equation}

\begin{equation}
   T_{c} =   T_{meson} = T_{QGP}.                                                         
\end{equation}

Analogous relations we have as well for baryons.

\section{Boltzmann equation for  the meson production}

Consider the process of meson production in QGP: 

\begin{equation}
q + \overline {q} \rightarrow  meson.
\end{equation}

This process is connected with the potential barrier corresponding to the confining potential for quarks in a spherical well.
Let  $ n(t,\boldsymbol {x})$  be a quark  density. 
We shall write the subscript  1 or 2 for $q$ and $\overline {q}$ respectively.

The general number of reactions (5)  per unit time  is equal to

\begin{equation}
            n_1(t,\boldsymbol {x})   n_2(t,\boldsymbol {x}) \int_{0}^{c}  \sigma (v) v  F(v) dv,                                        
\end{equation}
where $v$ is a relative velocity of particle, $c$ is the  velocity of light, $\sigma $ is the fusion cross-section that is determined  by $\sigma ={ \sigma}_0 P_{str} $, $F(v)$ is the distribution function  of relative velocity\cite {lif} \S 39. The quantity  $P_{str}$ is a probability of quark tunnelling to overcome the potential barrier, and $ {\sigma}_0 = \pi R_m^2$, where $R_m$ is a radius of meson sphere. According to (A.7) from Appendix A:

\begin{equation}
\sigma = \pi R_m^2  exp [- \frac {4 \sqrt {2MK_m}}{3\hbar} {R_m}^{3/2} ],
\end{equation}
where $K_m$  is a string tension for mesons,   $M$ is the reduced mass that is equal to 

\begin{equation}
M = \frac {m_1 m_2} {m_1 + m_2}.
\end{equation}

We presume that the spatial distribution of particles is homogeneous. Then $ n_1(t,\boldsymbol {x}) \equiv n(t)$ and $ n_2(t,\boldsymbol {x}) \equiv n(t)$.
We shall suppose as well that the  reverse process is absent.
We shall consider this suggestion  in the fifth section.

Finally, we can write  the Boltzmann equation

\begin{equation}
   \frac {dn}{dt}  = -   n^2(t)   Int ,                                                      
\end{equation}
where   the right part is called the collision integral, and  $Int$ is determined as follows:

\begin{equation}
    Int     = \sigma  \int_{0}^{c} v  F(v) dv.    
\end{equation}

The distribution $F(v)$  is taken   according to the expression:

\begin{equation}
    F(v) = 4 \pi \frac {M^3}{({2\pi M \kappa T})^{3/2}} v^2 exp ( - \frac  {Mv^2}{2 \kappa T} ).
\end{equation}

Using the relation

 \begin{equation}
    \int   x  e^{a x} dx = e^{a x} (\frac {x}{a} - \frac {1}{a^2})
 \end{equation}
and  denoting $\alpha = M/2 \kappa T$, we derive 

\begin{equation}
    I   =  \int_{0}^{c}   v^3  exp (- \alpha v^2 )dv  = -\frac {e^{-\alpha c^2}}{2}\bigg[\frac {c^2}{\alpha} + \frac {1}{{\alpha}^2}\bigg] +\frac {1}{2 {\alpha}^2}.
 \end{equation}

Hence, 

\begin{equation}
    Int =4 \sigma  \cdot  I \cdot \pi \frac {M^3}{({2\pi M \kappa T})^{3/2}}.  
\end{equation} 

To solve  Eq.(9), we use the separation of variables:

\begin{equation}
   \frac {dn}{n^2}  = -   dt \cdot  Int.                                                       
\end{equation}

From (15) we get

\begin{equation}
 -  \frac {1}{n}  = -   t  \cdot   Int     - C.                                                   
\end{equation}

A constant $C$ is determined from the initial condition $n(t=0) = n(0)$. 
Hence, we have

\begin{equation}
  \frac {1}{n}  =  t \cdot    Int  + \frac {1}{n(0)}.                                                     
\end{equation}

Finally, we obtain

\begin{equation}
  n(t)  =  \frac {n(0)} {Int \cdot   n(0) t  +1 }.                                              
\end{equation}

Substituting this expression in the right part of Eq.(9), we shall find the dependence of reaction rate  on  a  time, temperature, mass  of quark as well as 
on  a string tension for mesons and an initial value $n(0)$.

Using the expression (18), we can calculate the time of quark density reduction $\beta$ times:

\begin{equation}
 \tau (\beta)=  \frac {\beta -1} {Int \cdot   n(0) }.                                             
\end{equation}

The quantity $\beta= n(0)/n(\tau)$  points to  the degree of execution for the reaction (5) during the time $\tau$ . 

Let $n_{mes}$ be a meson density. Since

\begin{equation}
  \frac {d n_{mes}}{dt}  =  -  \frac {dn}{dt},                                                   
\end{equation}
and $ n_{mes} (t) + n(t) = const =n(0)$, then $n_{mes} = n(0) - n(t)$.

\section{Boltzmann equation for baryon production }
 
Consider the process of baryon production from QGP: 

\begin{equation}
 q_1 + q_2 \rightarrow  pair (q1, q2 ),                      
\end{equation}

\begin{equation}
     pair (q1, q2 )+q_3   \rightarrow  baryon.                    
\end{equation}

For quantities connected with the quarks $q_1, q_2, q_3$, and the pair $(q1, q2 )$ one enters the subscripts 1, 2, 3, and "pair" respectively.
Besides, for the free particles $q_1$ and $q_2$ we assume that $n_1(t) = n_2(t) = n(t)$.

Repeating the above  reasoning from the previous section,  we find that
\footnote {We suppose that $q_1$   and $q_2$ correspond to different  quark flavors. 
The case, under which $q_1$   and $q_2$  have the same flavor,  we shall consider in the next section.} 

\begin{equation}
   \frac {dn}{dt}  = -  n^2   Int 1,                                                       
\end{equation}
where  $Int1$ is determined as follows:

\begin{equation}
Int1 =   \int_{0}^{c}  \sigma_1 (v) v  F(v) dv. 
\end{equation}

If  the charges $e_1$ and $e_2$ are repulsed, we use Appendix A  \footnote {If the charges $e_1$ and $e_2$ are attracted, then  one needs to use the relations 
(13) and (14) from the previous section.}.
Since  the condition $M c^2/2 > e_ 1 e_2/R_{pair}$ usually  is  executed for the radius of the pair $(q1, q2 ) \, R_{pair}$, then the integral (24) is divided into two integrals:

\begin{equation}
	   Int1 = Ints+ Inti .
\end{equation}

The integral $Ints$ corresponds to the case when there are both the confining and Coulomb potential:

\begin{equation}
Ints =   \int_{0}^{v_0}  \sigma_1 v  F(v) dv,
\end{equation}
where $v_0$ is found from the condition:

\begin{equation}
	   \frac {e_1 e_2}{R_{pair}} = \frac {M v_0^2}{2}.
\end{equation}

Then according to (A.8) and (A.9) from Appendix A:

\begin{equation}
\sigma_1 = \pi R_{pair}^2   exp \bigg [- \frac {4 \sqrt {2M K_{pair}}}{3\hbar} R_{pair}^{3/2} \bigg] P_{coul},
\end{equation}
where $K_{pair}$ is a string tension for  the pair $(q1, q2 )$, and 

\begin{equation}
	  P_{coul} = exp \bigg [- \frac {2\sqrt {2M e_1 e_2 b}}{\hbar}\bigg (\frac {\pi}{2} -   2{ (\frac {R_{pair}}{b} )}^{1/2} + \frac {1}{3}{ (\frac {R_{pair}}{b} )}^{3/2}   \bigg) \bigg].   
\end{equation}

The quantity $b$ is determined from the condition

\begin{equation}
	   \frac {e_1 e_2}{b} = \frac {M v^2}{2}.
\end{equation}

The integral $Inti$ corresponds only to the confining potential:

\begin{equation}
Inti =   \int_{v_0}^{c}  \sigma \cdot v  F(v) dv,
\end{equation}
where  $\sigma$ is taken as in the previous section.

For the density $n(t)$ we have

\begin{equation}
 n(t) =\frac {n(0)}{Int1 \cdot t \cdot n(0) +1}.
\end{equation}

Consider now collisions of the pair $(q1, q2 )$   and $q_3$. According to the above reasoning

\begin{equation} 
   \frac {dn_3}{dt}  = -  n_{pair} n_3 Int 2.                                                       
\end{equation}

Since

\begin{equation} 
    \frac{d n_{pair}}{dt} =      \frac  {dn_3}{dt}     - \frac  {dn}{dt},                                                 
\end{equation}
then $ n_{pair}(t) + n(t) - n_3(t) = const$. We assume for simplicity  that $n_{pair}(0) + n(0) - n_3(0) = 0 $.
Hence, we get

\begin{equation} 
       \frac {dn_3}{dt}  = -  (n_3 - n) n_3   Int 2.                                            
\end{equation}

Eq.(35) is the special case of the Bernoulli equation (B.1) from Appendix B, where $y=n_3, x=t, h= Int2, a=Int1/Int2, b = 1/n(0)Int2$.
Using (B.9) from  Appendix B, we derive

\begin{equation} 
  n_3(t) = \frac {1}{{\bigg ( \frac{at+b}{b} \bigg )}^{-\frac {1}{a}} \bigg [ \frac  {1}{ n_3(0)}  +    \frac { h {(at+b)}^{\frac {a+1}{a}}}{(a+1) b^{1/a}} 
 - \frac {hb}{a+1}  \bigg ]}.
\end{equation}

Let $n_{bar}$ be a baryon density. Since

\begin{equation}
  \frac {d n_{bar}}{dt}  =  -  \frac {dn_3}{dt},                                                   
\end{equation}
and $ n_{bar} (t) + n_3(t) = const =n_3(0)$, then $n_{bar} = n_3(0) - n_3(t)$.

If  the charges $e_{pair}$ and $e_3$ are attracted, then the   integral $Int2$ is determined as follows:

\begin{equation}
Int2 = \sigma_2   \int_{0}^{c }  v  F(v) dv = 4 \sigma_2  \cdot  I \cdot \pi \frac {M_1^3}{({2\pi M_1 \kappa T})^{3/2}},
\end{equation}
where $M_1$ is the reduced mass 

\begin{equation}
M_1 = \frac {(m_1 + m_2) m_3} {m_1 + m_2 +m_3}.
\end{equation} 

We use $\sigma_2$ for the collision of the pair $(q1, q2 )$ and  $q_3$: 

\begin{equation}
\sigma_2 = \pi R_b^2  exp [- \frac {4 \sqrt {2M_1K_b}}{3 \hbar} {R_b}^{3/2} ], 
\end{equation}
where $R_b$   is a radius  of baryon,  and   $K_b$   is a string tension for baryons. 

Substituting (32) and (36)  to the right part of Eq.(35), we shall derive the dependence of reaction rate  on  a  time, temperature, mass and charge of quarks as well as  
on  a string tension and  initial values $n(0), n_3(0)$.

If  the charges $e_{pair}$ and $e_3$ are repulsed,  one needs to apply the above reasoning in this section based on Appendix A.

In conclusion, we note that one should be taken into account  the  permutations: $(q_1, q_2, q_3), (q_1, q_3, q_2), (q_3, q_2, q_1)$. 
Each permutation corresponds to a one-third of particles for each type.

\section{Collisions of ultrarelativistic heavy ions in colliders.}

According to modern representations the quark-gluon plasma can be formed in collisions of ultrarelativistic heavy ions.
Such plasma  can exist only under high enough temperatures  starting with 180-200  MeV  \cite { wong, kohs}.
Since produced hadrons  fly away or can decay  by  usual ways, reverse processes  of  hadron decay into quarks are absent for $T < 180\, MeV$.
This conclusion is based as well on the fact that the spectrum of produced particles does not contain quarks \cite {emel}, Ch.2.
This means that the probability of hadron decay into quarks  is very small.

First we shall consider the production of  mesons, which  is described by  Eq.(9).
We take as  examples of mesons: $J/\psi$ and ${\pi}^0$.
Parameters of these mesons are given in the table:

\begin{tabular}{lccr} 
\\
\hline
\hline
Type of meson    &   Structure         &  Mass of quark (MeV)   &   Time  $\tau(\beta=10^9) $   \, (s)        \\    
\hline
$J/\psi$             &  $c\overline{c}$  &  1300                         &   $1.98 \times 10^{-11}  $         \\
${\pi}^0$          &  $u\overline{u}$  &  5                             &   $2.63 \times 10^{-11}  $       \\   
\hline 
\hline
\\
\end{tabular}

To calculate the  integral $Int$, we use the expressions (7) and (14), where  $K_m=0.9 \, GeV {fm}^{-1} = 1.44 \times 10^{10}\, erg/cm, R_m = 1 \, fm = 10^{-13} \, cm,  T = 170 \, MeV = 2.0 \times 10^{12} \,K$.
The initial quark density is determined according to (2).
We have $n(0) = 3 g \zeta(3) T^3/4\pi^2 \approx 0.8 \times 10^6 \, MeV^2  \approx 1.8 \times 10^{38} \, cm^{-3}$.
Next we compute $\tau (\beta)$ by means of the expression (19).

Now we can pass on to the baryon production.
We take as  examples of baryons: $\Sigma^+$ and $p$.
Parameters of these baryons are given in the table: 

\begin {tabular}{lccr} 
\\
\hline
\hline
Type of baryon    & Structure    & Masses of quarks (MeV)    &  Charges   of quarks  \\
\hline
$\Sigma^+$      & $uus$        &  5, 5, 150                       &  2/3, 2/3, -1/3         \\
$p$                   & $uud$        &  5, 5, 7                          &  2/3, 2/3, -1/3      \\
\hline
\hline 
 \\
\end {tabular}

We consider the reactions:

\begin{equation}
 u + u \rightarrow  pair (u, u),                      
\end{equation}

\begin{equation}
     pair (u, u )+ d(s)   \rightarrow  p (\Sigma^+).                    
\end{equation}

Let $n(t)$ be the density of the quark $u$.
Then we have

\begin{equation} 
    \frac{d n}{dt} =    2 Int1 \frac {n(n-1)} {2} \approx     Int1 \cdot n^2                                             
\end{equation}

The solution of this equation is

\begin{equation}
 n(t) =\frac {n(0)}{Int1 \cdot t \cdot n(0) +1}.
\end{equation}

Consider now collisions of the pair $(u, u)$   and $ p (\Sigma^+)$. According to the above reasoning

\begin{equation} 
   \frac {dn_3}{dt}  = -  n_{pair} n_3 Int 2.                                                       
\end{equation}

Since

\begin{equation} 
    \frac{d n_{pair}}{dt} =      \frac  {dn_3}{dt}  -  \frac  {1}{2}  \frac  {dn}{dt},                                                 
\end{equation}
then $ n_{pair}(t) +  \frac  { n(t)}{2} - n_3(t) = const$.
 For simplicity, we suppose that $n_{pair}(0) +  \frac  {n(0)}{2}  - n_3(0) = 0 $.
Hence, we shall derive

\begin{equation} 
       \frac {dn_3}{dt}  = -  (n_3 - \frac  {n}{2} ) n_3   Int 2 = - {n_3}^2  Int 2    + \frac {n(0) Int2\cdot n_3} {2 Int1 \cdot n(0)\, t +2}                                        
\end{equation}

Eq.(47) is the special case of the Bernoulli equation  (B.1) from Appendix B, where $y=n_3, x=t, h= Int2, a=2 Int1/Int2, b = 2/n(0)Int2$.
Using (B.9), we write

\begin{equation} 
  n_3(t) = \frac {1}{{\bigg ( \frac{at+b}{b} \bigg )}^{-\frac {1}{a}} \bigg [ \frac  {1}{ n_3(0)}  +    \frac { h {(at+b)}^{\frac {a+1}{a}}}{(a+1) b^{1/a}} 
 - \frac {hb}{a+1}  \bigg ]}.
\end{equation}

By means of the expression
\begin{equation}
   \frac { n(0)}{n_3(\tau)} = \beta(\tau) ={\bigg[  \frac {at+b} {b}  \bigg]}^{-1/a}   +  n_3(0)\frac { h (at+b)}{a+1} - n_3(0) {\bigg[  \frac {at+b} {b}  \bigg]}^{-1/a} 
\frac {hb}{a+1},
\end{equation} 
we can find the functional dependence $\beta(\tau)$.

For $\Sigma^+ $ with  $\tau = 10^{-11} s$   we get $\beta =5.3 \times 10^{12} $.   
For $p $ with  $\tau = 10^{-11} s$   we have $\beta =  4.92 \times 10^{12} $.  
Since there are only two variants of reactions, we take $n(0) = 0.9 \times 10^{38} cm^{-3}$ and  $n_3(0) = 0.45 \times 10^{38} cm^{-3}$. 

\section{Phase transition of QGP-hadrons in the early Universe}

In modern cosmology the central meaning has the process of inflation, which is explained by means of existence of a scalar field \cite {gorb}.
At the end of inflation  this scalar field decays and generates elementary particles: electrons, positrons, quarks, antiquarks, photons, gluons, etc \cite {gorb}.
Next there is the thermodynamic equilibrium between QGP and hadron gas for the temperature of order 200 MeV \cite { kap, kohs}.
When the temperature decreases, one can disregard  processes of hadron decay into quarks using results of  ultrarelativistic heavy ion  collisions  in collider.
Finally, quarks and antiquarks vanish in the free state.

First we consider the production of  mesons.
One should be modified the kinetic equation (9) taking into account expanding  Universe \cite {rub}, Ch.9.
We have

\begin{equation}
   \frac {dn}{dt}  + H n +   n^2  Int   = 0 .                                                    
\end{equation}

The quantity $H$ is the Hubble parameter:
\begin{equation}
   H(t_c) = \frac {\dot {a}(t_c)}{a(t_c)},                                                     
\end{equation}
where $a(t_c)$ is the scale factor \cite {rub} , Sec.1.2, and the dot denotes the derivative with respect to the cosmic time $t_c$.

The time $\theta$ for the phase transition from quark-gluon plasma to hadron matter in the early Universe is approximately equal to $10^{-5} \, s$.
We shall apply the subscript 0 for the present time. 
To compute $H (t_c)$, we use the approximate expression (see \cite{rub}, Sec 4.5):

\begin{equation}
   H^2 (t_c) = H^2_0 \bigg [\Omega_M (\frac {a_0}{a (t_c)})^3  +   \Omega_{\Lambda}\bigg ],                                         
\end{equation}
where $\Omega_M \approx 0.315,  \Omega_{\Lambda} \approx 0.685, H_0 \approx 67 \, km/s \cdot Mpc \approx 0.21 \times 10^{-17}\, s^{-1}$. 

The solution of Eq. (52) is

\begin{equation}
   a (t_c) = a_0 {( \frac {\Omega_M }{ \Omega_{\Lambda}})}^{1/3} { \bigg[ sh (\frac {3}{2}\sqrt { \Omega_{\Lambda}} H_0 t_c) \bigg ] }^{2/3}                                             
\end{equation}

Substituting this expression for $\theta=10^{-5} \, s$ to Eq. (52), we obtain $H(\theta) = 6.67 \times 10^4 \, s^{-1}$.

Eq. (50) is the Bernoulli equation (B.10) from Appendix B, where $y=n, x=t, a= H(\theta), b = Int.$
Using (B.13), we shall derive the solution of  Eq. (50):

\begin{equation} 
  n (t ) =\frac {n(0)} { e^{H (\theta) t} - n(0) \frac {Int}{H(\theta)} (1 - e^{H (\theta) t})}.
\end{equation}

Then

\begin{equation} 
  \beta (\tau) = \frac  {n(0)}{n(\tau)} =     { e^{H(\theta) \tau} - n(0) \frac {Int}{H(\theta)} (1 - e^{H(\theta) \tau})}.
\end{equation}

Since  $H(\theta)\tau << 1$ and

\begin{equation} 
  \frac {1 - e^{H(\theta) \tau} }{H(\theta)} \approx -\tau,
\end{equation}

 we have

\begin{equation} 
  \beta (\tau) =    1 + n(0) Int \cdot \tau.
\end{equation}

Consequently, we have obtained the same expression (19), i.e.,  expanding  does not influence on kinetic processes in QGP.
Hence, we can use all results of the previous section derived for collisions of ultrarelativistic heavy ions.
Such conclusion is true as well for baryons. 

\section{Conclusion}

Modern cosmology supposes that the quark-gluon plasma(QGP)  arises after the inflation as a result of inflaton decay.
On the other hand such state of matter is observed in ultrarelativistic heavy ion collisions.
The quark-gluon plasma generates baryons and mesons by means of  quark fusion reaction.
Investigating these reactions is one of the actual problems for QGP theory.
The hadron production depends on a quark density, temperature, relative velocity under collisions, etc.
Moreover particles  must overcome a potential barrier connected with  the quark  confinement.
Using the method of velocity averaging and  quasi-classical approach of quantum mechanics it can formulate nonlinear kinetic equations.
The solutions of these equations permit to find the  reaction rate  as a function of a  time, temperature, quark mass (charge), and a  string tension connected
 with the confining potential. 
One should be noted that the method of velocity averaging can be used as well for processes of quark annihilation and usual hadron decay in QGP.
The proposed form of the interaction potential between quarks gives a new approach to investigate phase transitions in QGP. 

\appendix

\section {Tunnelling for baryon and meson}

To study tunnelling, we use the quasi-classical approximation  in quantum mechanics \cite {land}, Ch. VII.
Based on the  spherical  model with the radius $R$, we suppose the next interaction potential between quarks $V(r)$ (see Fig.1 a)).

\begin{itemize}
\item In the interior of the sphere the confining potential is equal to   $V(r) = K r $ according to the string model \cite {wong}, Sec.15.
The quantity $K$ is called a string tension.
\item On the boundary of the sphere  there is a jump.
\item On the outside the potential  is given in terms of
\begin{equation}
          V(r) = -   \frac {4\alpha_{s} }{3 r} exp (- \frac {r}{\lambda}).                                        
\end{equation}
This expression corresponds to  QGP screening \cite {wong}, Sec.15, where $\alpha_s$ is the coupling constant, $\lambda$ is the characteristic distance for QGP.
In the case of repulsion for quarks one should be used the repulsing Coulomb potential instead the screening potential (see Fig.1 b)).
\end{itemize}

The kinetic energy of quark at infinity  $E_{kinetic}$  is  equal to $mv^2/2$ for the relative velocity $v$. Then the turning point  $a$ is determined from the condition:
 
\begin{equation}
          \frac {mv^2}{2} = K a.                                          
\end{equation}
 
We have a simple approximate expression for the  tunnelling  probability for some value of $E_{kinetic}$ and the corresponding value of $a$  \cite {land}, Ch. VII.:

\begin{equation}
	         P_{str} = exp \bigg[- \frac {2}{\hbar} \int_{a}^{R} \sqrt {2 m ( K r -K a)} \, dr   \bigg] .
\end{equation}

To compute the integral in (A.3), we use the substitution $r - a = x$. Then 
\begin{equation}
	          \int_{a}^{R} \sqrt {r-a } dr = \int_{0}^{R-a} \sqrt {x } dx =\frac {2}{3} {(R-a)}^{3/2}.
\end{equation}

\includegraphics{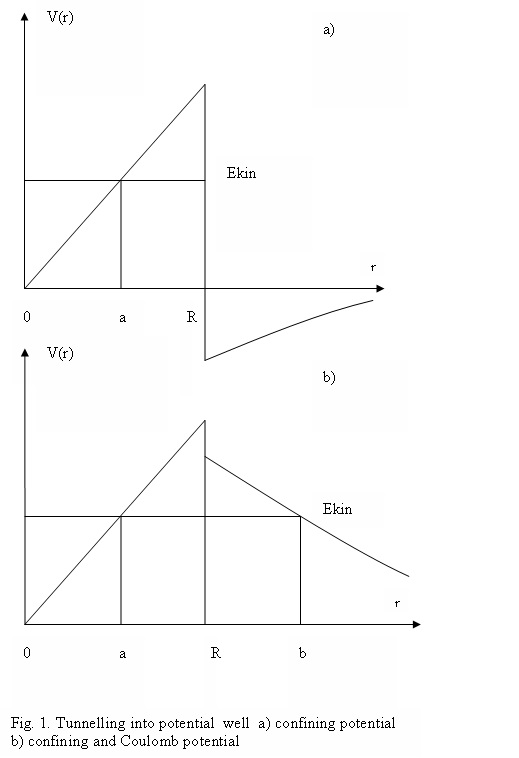}
  
As a result,  we have the  simple formula for the tunnelling probability:

\begin{equation}
	  P_{str}  = exp \bigg[- \frac {4 \sqrt {2mK}}{3\hbar} {(R-a)}^{3/2}\bigg].
\end{equation} 
Finally, we get

\begin{equation}
	  P_{str}  = exp \bigg [- \frac {4 \sqrt {2mK}}{3\hbar} {(R-\frac {mv^2}{2K})}^{3/2} \bigg].    
\end{equation}

For the problem under consideration $ R K>> mv^2/2$. 
In this case
\begin{equation}
	  P_{str}  = exp \bigg [- \frac {4 \sqrt {2mK}}{3\hbar} R^{3/2} \bigg] .   
\end{equation}

Consider now the case with the repulsing Coulomb potential.
Let $e_1$ and $e_2$ be the charges of repulsing quarks.
Then there are both the confining and Coulomb potential. The probability of fusion is

\begin{equation}
P_{\Sigma}  = P_{str}   \cdot  P_{coul} =  exp [- \frac {4 \sqrt {2m K}}{3\hbar} R^{3/2} ] P_{coul},
\end{equation}
where $P_{coul}$ is connected with the Coulomb barrier.

According to \cite {land}, Ch. VII one needs to calculate the integral

\begin{equation}
	  P_{coul} = exp \bigg [- \frac {2\sqrt {2m}}{\hbar}\int_{R}^{b} \sqrt {(\frac {e_1 e_2}{r} -  \frac {e_1 e_2}{b})} \, dr \bigg],   
\end{equation}
where $b$ is determined by the condition

\begin{equation}
	   \frac {e_1 e_2}{b} = \frac {mv^2}{2}.
\end{equation}

The integral  in (A.9) is computed as follows:

\begin{equation}
	 \int_{R}^{b} \sqrt {(\frac {1}{r} -  \frac {1}{b})} \, dr =    \int_{0}^{b} \sqrt {(\frac {1}{r} -  \frac {1}{b})} \, dr -  
\int_{0}^{R} \sqrt {(\frac {1}{r} -  \frac {1}{b})}\, dr.
\end{equation}

To find the first integral in the right part of (A.11 ), we enter the new variable $x = r/b$. Then

\begin{equation}
 \int_{0}^{b} \sqrt {(\frac {1}{r} -  \frac {1}{b})} \, dr = \sqrt {b} \int_{0}^{1} \sqrt {(\frac {1}{x} -  1)}\, dx.
\end{equation}

This integral is calculated by the substitution $x=sin^2\theta$:

\begin{equation}
  \int_{0}^{1} \sqrt {(\frac {1}{x} -  1)}\, dx = \frac {\pi}{2}.
\end{equation}

To find the second integral in the right part of (A.11), we use the approximate expression for $x<<1$:

\begin{equation}
\sqrt { (\frac {1}{x} -  1)}  \approx x^{-1/2} -\frac {x^{1/2}}{2}.
\end{equation}

Then

\begin{eqnarray}
  \int_{0}^{R} \sqrt {(\frac {1}{r} -  \frac {1}{b})}\, dr = \sqrt {b}\int_{0}^{R/b} \sqrt {(\frac {1}{x} -  1)} \,dx 
\approx \sqrt {b}\bigg (2{ (\frac {R}{b} )}^{1/2} - \frac {1}{3}{ (\frac {R}{b} )}^{3/2}\bigg).    \nonumber \\
\end{eqnarray}

Finally, we have

\begin{equation}
	  P_{coul} \approx exp \bigg [- \frac {2\sqrt {2m e_1 e_2 b}}{\hbar}\bigg (\frac {\pi}{2} -   2{ (\frac {R}{b} )}^{1/2} +
 \frac {1}{3}{ (\frac {R}{b} )}^{3/2}   \bigg) \bigg]   
\end{equation}

\section {Bernoulli equation}

Consider the special case of the Bernoulli equation:

\begin{equation} 
       \frac {dy}{dx}  = -  h y^2 + \frac {y}{a x +b}.                                             
\end{equation}

This equation is reduced to the linear equation by the substitution $u = y^{-1}$.  
Then we get

\begin{equation} 
       \frac {du}{dx}  + \frac {u}{a x +b}   - h = 0.                                       
\end{equation}

Introduce the notations $P(x) =  \frac {1}{a x +b}$ and  $Q(x) =  -h $.
According to the theory of the ordinary differential equation of the first order \cite {kam}, Ch.1:

\begin{equation} 
       u = e^{- \int_{0}^{x} P(x) dx}   \bigg [   C -   \int_{0}^{x} Q(x)  e^{ \int_{0}^{x} P(x) dx } dx \bigg]. 
\end{equation}      

Since

\begin{equation} 
        \int_{0}^{x} P(x) dx =   \int_{0}^{x} \frac {dx}{ax+b} = \frac {1}{a} ln (\frac {ax+b}{b})
\end{equation} 
then
\begin{equation} 
      e^{ \int_{0}^{x} P(x) dx} =     {\bigg ( \frac{ax+b}{b} \bigg )}^{\frac {1}{a}},
\end{equation} 

\begin{equation} 
       e^{ -\int_{0}^{x} P(x) dx} =   {\bigg ( \frac{ax+b}{b} \bigg )}^{-\frac {1}{a}},
\end{equation} 
and

\begin{equation} 
  \int_{0}^{x} Q(x)  e^{ \int_{0}^{x} P(x) dx } dx = - \int_{0}^{x}   h  {\bigg ( \frac{ax+b}{b} \bigg )}^{\frac {1}{a}} dx = 
 - \frac { h {(ax+b)}^{\frac {a+1}{a}}}{(a+1) b^{1/a}} + \frac {hb}{a+1}.
\end{equation}

Hence,

\begin{equation} 
  u = {\bigg ( \frac{ax+b}{b} \bigg )}^{-\frac {1}{a}} \bigg [ u(0)   +    \frac { h {(ax+b)}^{\frac {a+1}{a}}}{(a+1) b^{1/a}}  - \frac {hb}{a+1}  \bigg ].
\end{equation}

Since $y=1/u$ and $y(0)=1/u (0)$, we derive

\begin{equation} 
  y = \frac {1}{{\bigg ( \frac{ax+b}{b} \bigg )}^{-\frac {1}{a}} \bigg [   \frac {1}{y(0)}  +    \frac { h {(ax+b)}^{\frac {a+1}{a}}}{(a+1) b^{1/a}}  - \frac {hb}{a+1}  \bigg ]}.
\end{equation}

Moreover, one needs to consider the Bernoulli equation

\begin{equation} 
       \frac {dy}{dx}  + a y + b y^2 = 0,                                           
\end{equation}
which is reduced as well  by the substitution $u = y^{-1}$  to the linear equation

\begin{equation} 
       \frac {du}{dx} - au  - b = 0.                                        
\end{equation}

Enter the notations $P(x) =  -a$ and  $Q(x) =  - b $.
Using the expression (B.3), we derive

\begin{equation} 
  u = C e^{ax} - \frac {b}{a} (1 - e^{ax}).
\end{equation}

Finally, we have

\begin{equation} 
  y =\frac {y(0)} { e^{ax} - y(0) \frac {b}{a} (1 - e^{ax})}.
\end{equation}

\end{document}